\documentclass[a4paper]{article}

\usepackage{INTERSPEECH2020}
\usepackage{multirow}
\usepackage[utf8]{inputenc}
\usepackage{xcolor}
\usepackage{ulem}
\usepackage{hyperref}

\title{Multilingual Bottleneck Features for Improving ASR Performance of Code-Switched Speech in Under-Resourced Languages}
\name{Trideba Padhi, Astik Biswas, Febe de Wet, Ewald van der Westhuizen \& Thomas Niesler}

\address{
Department of Electrical and Electronic Engineering, Stellenbosch University\\
Stellenbosch, South Africa
}

\email{\{tpadhi, abiswas, fdw, ewaldvdw \& trn\}@sun.ac.za}

\begin{document}

\maketitle

\begin{abstract}
In this work, we explore the benefits of using multilingual bottleneck features (mBNF) in acoustic modelling for the automatic speech recognition of code-switched (CS) speech in African languages. 
The unavailability of annotated corpora in the languages of interest has always been a primary challenge when developing speech recognition systems for this severely under-resourced type of speech. 
Hence, it is worthwhile to investigate the potential of using speech corpora available for other better-resourced languages to improve speech recognition performance. 
To achieve this, we train a mBNF extractor using nine Southern Bantu languages that form part of the freely-available multilingual NCHLT corpus. 
We append these mBNFs to the existing MFCCs, pitch features and i-vectors to train acoustic models for automatic speech recognition (ASR) in the target code-switched languages.
Our results show that the inclusion of the mBNF features leads to clear performance improvements over a baseline trained without the mBNFs for code-switched English-isiZulu, English-isiXhosa, English-Sesotho and English-Setswana speech. 

This represents a step forward in the use of out-of-domain data to improve the automatic recognition of code-switched speech in under-resourced South African languages.
\end{abstract}


\section{Introduction}
\label{sec1.0_intro}
With recent rapid advances in the field of artificial intelligence, the ease with which humans can interact with machines has become a yardstick with which the sophistication of a system is assessed~\cite{hinton2012deep}. 
This has had a particularly strong effect in stimulating research interest in ASR. 
However almost all current speech interfaces assume monolingual input, while most of the world's population is conversant in more than one language.
Hence, there had recently also been a surge in interest in the automatic recognition of code-switched speech. 

The population of South Africa is highly multilingual and this has recently motivated the development of code-switching ASR systems for African languages~\cite{biswas2018multilingual,yilmaz2018building}. 
In South Africa, most code-switching occurs between English and one or more South African languages.
However, annotated speech corpora that include such mixed-language speech are  extremely scarce and those that are available are small. 
Several approaches have been proposed to address the limitations posed by this lack of annotated speech data.
One major drive considers the incorporation of speech in other better-resourced languages to leverage improved ASR performance in the target languages. 
In \cite{vu2012multilingual}, the authors show that the overall performance of a multilayer perceptron acoustic model increases substantially when the system is initialized using bottleneck features (BNFs). 
This acoustic modelling strategy was coupled with a new language modelling strategy called ``open target language'' which trains more flexible models for language adaptation and with which improvements in performance were reported for under-resourced languages.
In \cite{matejka2014neural}, improvements in the region of 45\% over baseline features were reported when incorporating BNFs for ASR on DARPA RATS data.
Two deep bottleneck neural networks were trained on English and Mandarin and the resulting features fused in~ \cite{jiang2014deep}, yielding improvements of between 2\% and 7\% in equal error rate for longer and shorter segments respectively on the NIST language recognition evaluation 2009 (LRE09) dataset.
The authors of \cite{fer2017multilingually} analysed the practical aspects of training bottleneck networks as well as their integration in ASR. 
They also compared monolingual and multilingual training for ASR by evaluating different systems on the LRE09 dataset.

A BNF extractor that was specifically designed for subword modelling and was trained on the GlobalPhone database was proposed in \cite{hermann2018multilingual}. 
From 16 of the languages in the Globalphone corpus, 10 high resource languages were used for training the extractor and the remaining 6 for ASR performance evaluation.
It was shown that an ASR system trained on a single language using this BNF extractor outperformed a baseline whose features were computed using a correspondence autoencoder and vocal tract length normalization. It was also found that using two or more languages in the BNF extractor training pool resulted in better performance than using a training data set of the same size from only one language.

In previous work we have explored the effectiveness of using out-of-domain monolingual South African speech to improve the performance of code-switched ASR~\cite{biswas2018improving}.
We found that better-resourced monolingual speech helped to enhance code-switched ASR performance, but only by a small margin considering the amount of out-of-domain data and the computational resources that were required to incorporate the additional data during acoustic model training.
Much more out-of-domain data than in-domain data was required to improve code-switched speech recognition accuracy, and thus the most effective way of improving performance has remained the extension of the in-domain training set.

However, given the severe scarcity of resources in the target languages,  we have also actively explored other ways to exploit available sources of out-of-domain data.
This study represents our first use of BNF extractors to leverage out-of-domain data to improve the accuracy of code-switched speech in five South African languages. Although BNF extractors are generally well established in ASR and other speech processing tasks, to the best of our knowledge, this is the first attempt to train BNF extractors using South African Bantu languages.

We investigate the benefit of training a BNF feature extraction network on related but out out-of-domain data and then using the extracted BNF features in combination with baseline features (MFCC, pitch and i-vectors) to train acoustic models for South African code-switch ASR. 
To achieve this, a multilingual BNF (mBNF) extractor is developed using nine South African Bantu languages from the freely-available multilingual NCHLT corpora~\cite{barnard2014nchlt}. Two mBNF extractors with different bottleneck dimensions are trained and used to extract BNFs from the target code-switched speech.
We observe that the incorporation of the mBNFs improves the code-switched speech recognition accuracy relative to the system trained using the baseline features.

\vspace{-1mm}
\section{Data}
\label{sec2.0_data}
This section introduces two data sets: a set of monolingual corpora in South Africa's 11 official languages that was used to train our mBNF extractor and a corpus of code-switched (CS) South African speech that was used to train acoustic models for ASR purposes.

\subsection{Monolingual Speech Data}
\label{sec2.1}
The NCHLT speech corpora contain monolingual wide-band prompted speech in each of the eleven official languages of South Africa~\cite{barnard2014nchlt}. 
A greedy algorithm was used to select the prompts from a body of text during the compilation of each corpus~\cite{eiselen2014developing}. Trigram or five-gram prompts were derived from the text data, depending on the orthographic conventions of each language.
This approach resulted in prompts that vary in length from single word utterances to short phrases of up to 10 words.
The South African English and Afrikaans corpora were not included in our current investigation, only data from the nine remaining languages that all belong to the Bantu language family were used. 

\begin{table}[h]
\footnotesize
\caption{Statistics of the training sets of the NCHLT Bantu speech corpora.}
\begin{tabular*}{\columnwidth}{@{\extracolsep{\fill}} l @{} r @{} r @{} r @{} r @{}}
\toprule
\multirow{2}{*}{\textbf{Language}} & \multirow{2}{*}{\textbf{Speakers}} & \textbf{Duration} & \textbf{Word} & \textbf{Word\phantom{0}} \\
  &   & \textbf{(hours)} & \textbf{types} & \textbf{tokens} \\
\midrule
IsiNdebele (nbl)  & 132                                  & 47.3                                        & 14\,679                             & 132\,529                             \\
Sepedi (nso)      & 194                                  & 50.6                                        & 11\,056                             & 266\,859                             \\
Sesotho (sot)     & 194                                  & 50.7                                        & 10\,424                             & 250\,125                             \\
SiSwati (ssw)     & 181                                  & 48.4                                        & 11\,925                             & 115\,611                             \\
Setswana (tsn)    & 194                                  & 50.6                                        & 5\,495                              & 254\,274                             \\
Xitsonga (tso)    & 182                                  & 49.6                                        & 5\,934                              & 208\,684                             \\
Tshivenda (ven)   & 192                                  & 49.3                                        & 7\,579                              & 218\,820                             \\
IsiXhosa (xho)    & 193                                  & 49.1                                        & 27\,856                             & 122\,236                             \\
IsiZulu (zul)     & 194                                  & 48.3                                        & 23\,912                             & 116\,319                             \\
\midrule
\textbf{Total}    & \textbf{1\,656}                      & \textbf{443.9}                              & \textbf{118\,860}                   & \textbf{1\,685\,457}                 \\
\bottomrule
\end{tabular*}
\label{tab:NCHLT-stat}
\vspace{-3mm}
\end{table}

We used the predefined NCHLT training sets\footnote{The definitions of the predefined NCHLT training, development and test sets are available at \url{https://sites.google.com/site/nchltspeechcorpus/}} summarised in Table~\ref{tab:NCHLT-stat} to train the feature extraction network introduced in Section~\ref{sec:mBNFex}.
The NCHLT development and test sets were not used.

\subsection{Code-switched Speech Data}
\label{SEC:corpus}

For building code-switched (CS) ASR systems, a dataset of multilingual speech was compiled from South African soap opera episodes~\cite{van2018city}.
The data contains examples of code-switching between four language pairs: English-isiZulu (EZ), English-isiXhosa (EX), English-Setswana (ET) and English-Sesotho (ES). 
IsiZulu and isiXhosa belong to the Nguni language family whereas Setswana and Sesotho belong to the Sotho-Tswana family.
Both these belong to the larger Southern Bantu language family. The available training data consists of three subsets: (1) manually segmented and transcribed data; (2) manually segmented but automatically transcribed data; and (3) automatically segmented and transcribed data.
The subsets are described in more detail below. When combined, the the three subsets contain 78.1 hours of speech. The development and test sets were taken from the manually segmented and transcribed data.

\subsubsection{Manually segmented and transcribed data}
In most of our experiments concerning code-switched speech, we have used 
a 23-hour set of of annotated speech.
This set was partitioned into a training set of 21.1 hours and development and test sets of 48.3 and 78 minutes respectively. The training set includes a language balanced subset as well as additional data that, although skewing the data towards English, was found to enhance ASR performance~\cite{biswas2019EZsemi}.
Table~\ref{tab:manual_train} gives an overview of the manually transcribed component of the training data used in this study.
 
\begin{table}[h]
\caption{Duration in minutes (m) and hours (h) as well as word type and token counts for the unbalanced manually segmented and transcribed training set.}
\centering 
\footnotesize
\begin{tabular*}{\columnwidth}{@{\extracolsep{\fill}} l r @{\hspace{1em}} r @{\hspace{1em}} r @{\hspace{1em}} r @{\hspace{1em}} r @{\hspace{1em}} r @{}}
\toprule
{\bf{Language}} & \begin{tabular}[c]{@{}c@{}}\bf{Mono}\\  (m)\end{tabular} & \begin{tabular}[c]{@{}c@{}}\bf{CS}\\ (m)\end{tabular}&
\begin{tabular}[c]{@{}c@{}}\bf{Total}\\ (h)\end{tabular}&
\begin{tabular}[c]{@{}c@{}}\bf{Total}\\ (\%)\end{tabular}&
\begin{tabular}[c]{@{}c@{}}\textbf{Word}\\ \textbf{tokens}\end{tabular} & \begin{tabular}[c]{@{}c@{}}\textbf{Word} \\ \textbf{types}\end{tabular} \\
\midrule
English & 755.0              & 121.8  & 14.6 & 69.3 & 194\,426              & 7\,908              \\
isiZulu & 92.8               & 57.4  & 2.5 & 11.9 & 24\,412              & 6\,789              \\
isiXhosa & 65.1               & 23.8  & 1.5 & 7.0  & 13\,825              & 5\,630              \\ 
Setswana & 36.9               & 34.5  & 1.2 & 5.6  & 21\,409              & 1\,525              \\
Sesotho & 44.7               & 34.0  & 1.3 & 6.2  & 22\,226              & 2\,321              \\
\midrule
{\bf{Total}} & 994.5              & 271.5  & 21.1 & 100.0  & 276\,290              & 24\,170              \\ 
\bottomrule
\end{tabular*}
\label{tab:manual_train}
\end{table}

A similar overview of the development and test sets is given in Table~\ref{tab:manual_dev_tst}.
It is noteworthy that the test sets present a strict evaluation as these utterances are never monolingual but always contain code-switching.

\begin{table}[h]
\caption{Duration in minutes of English, isiZulu, isiXhosa, Sesotho and Setswana monolingual (mdur) and code-switched (cdur) segments in the development and test sets.}
\footnotesize
	\centering
		\begin{tabular*}{\columnwidth}{@{\extracolsep{\fill}}c r r r r r @{}}
			\toprule
			\multicolumn{6}{c}{\textbf{English-isiZulu}} \\
			& emdur & zmdur & ecdur & zcdur & \textbf{Total} \\
			\textbf{Dev} & 0.0 & 0.0 & 4.0 & 4.0 & 8.0 \\
			\textbf{Test} & 0.0 & 0.0 & 12.8 & 17.9 & 30.4 \\ \midrule
			\multicolumn{6}{c}{\textbf{English-isiXhosa}} \\
			& emdur & xmdur & ecdur & xcdur & \textbf{Total} \\
			\textbf{Dev} & 2.9 & 6.5 & 2.2 & 2.1 & 13.7 \\
			\textbf{Test} & 0.0 & 0.0 & 5.6 & 8.8 & 14.3 \\ \midrule
			\multicolumn{6}{c}{\textbf{English-Setswana}} \\
			 & emdur & tmdur & ecdur & tcdur & \textbf{Total} \\
			\textbf{Dev} & 0.8 & 4.3 & 4.5 & 4.3 & 13.8 \\
			\textbf{Test} & 0.0 & 0.0 & 8.9 & 9.0 & 17.8 \\ \midrule
			\multicolumn{6}{c}{\textbf{English-Sesotho}} \\
			 & emdur & smdur & ecdur & scdur & \textbf{Total} \\
			\textbf{Dev} & 1.1 & 5.1 & 3.0 & 3.6 & 12.8 \\
			\textbf{Test} & 0.0 & 0.0 & 7.8 & 7.7 & 15.5 \\ \bottomrule
		\end{tabular*}
		\vspace{-1mm}
	\label{tab:manual_dev_tst}
\end{table}

\subsubsection{Manually segmented and automatically transcribed data}
During the compilation of the corpus described in the previous section, some data was manually segmented but not transcribed. 
In a previous investigation, these segments were transcribed using a semi-supervised procedure, resulting in an additional 11 hours of training data~\cite{biswas2019EZsemi}.

\subsubsection{Automatically segmented and transcribed data}

The semi-supervised approach applied in the previous section was subsequently extended to include a CNN-GMM-HMM based VAD system (without speaker diarization). 
This system was used to segment the raw audio of additional soap opera episodes and the resulting segments were transcribed in a semi-supervised manner~\cite{biswas2020SLTU}. Using this procedure a further 45.6 hours of training data was generated. Table~\ref{tab:semis_segs} provides a summary of the language tags assigned to the data by the semi-supervised procedures.

\begin{table}[h]
\caption{Number of segments assigned to each language by the semi-supervised transcription systems.}
\centering 
\footnotesize
\begin{tabular*}
{\columnwidth}{@{\extracolsep{\fill}} l r @{\hspace{1em}} r @{\hspace{1em}} r @{\hspace{1em}} r @{\hspace{1em}} r @{\hspace{1em}} r @{}}
\toprule
{\bf{Language}} & {\bf{Eng}} & {\bf{Zul}} & {\bf{Xho}} & {\bf{Sot}} & {\bf{Tsn}} & {\bf{CS}}
\\ \midrule
Man Seg & 2~780 & 3~113 & 657 & 3~370 & 32 & 13~338 \\
Auto Seg & 4~754 & 2~122 & 236 & 719 & 2~196 &  17~911\\
\bottomrule
\end{tabular*}
\label{tab:semis_segs}
\vspace{-3mm}
\end{table}

\section{Multilingual Bottleneck Feature Extraction}
\label{sec:mBNFex}
Multilingual bottleneck features (mBNF) have been shown to outperform traditional spectral features as well as monolingual bottleneck features in a variety of speech processing tasks \cite{fer2017multilingually,hermann2018multilingual,vesely2012language}.
Table~\ref{tab:Extractors} provides an overview of two mBNF extractors that were evaluated in this study. Details on each configuration are provided in subsequent sections.

\begin{table}[ht]
\caption{BNF extractor configurations.}
\footnotesize
\begin{tabular*}{\columnwidth}{@{\extracolsep{\fill}} l c c @{}}
\toprule
\textbf{Extractor} & \textbf{Training data} & \textbf{BNF dimension} \\ \midrule
mBNF$_1$ & NCHLT &  39 \\
mBNF$_2$ & NCHLT &  80 \\
\bottomrule
\end{tabular*}
\label{tab:Extractors}
\vspace{-2mm}
\end{table}

\begin{figure}[t]
	\centering
    \includegraphics[width=0.9\columnwidth]{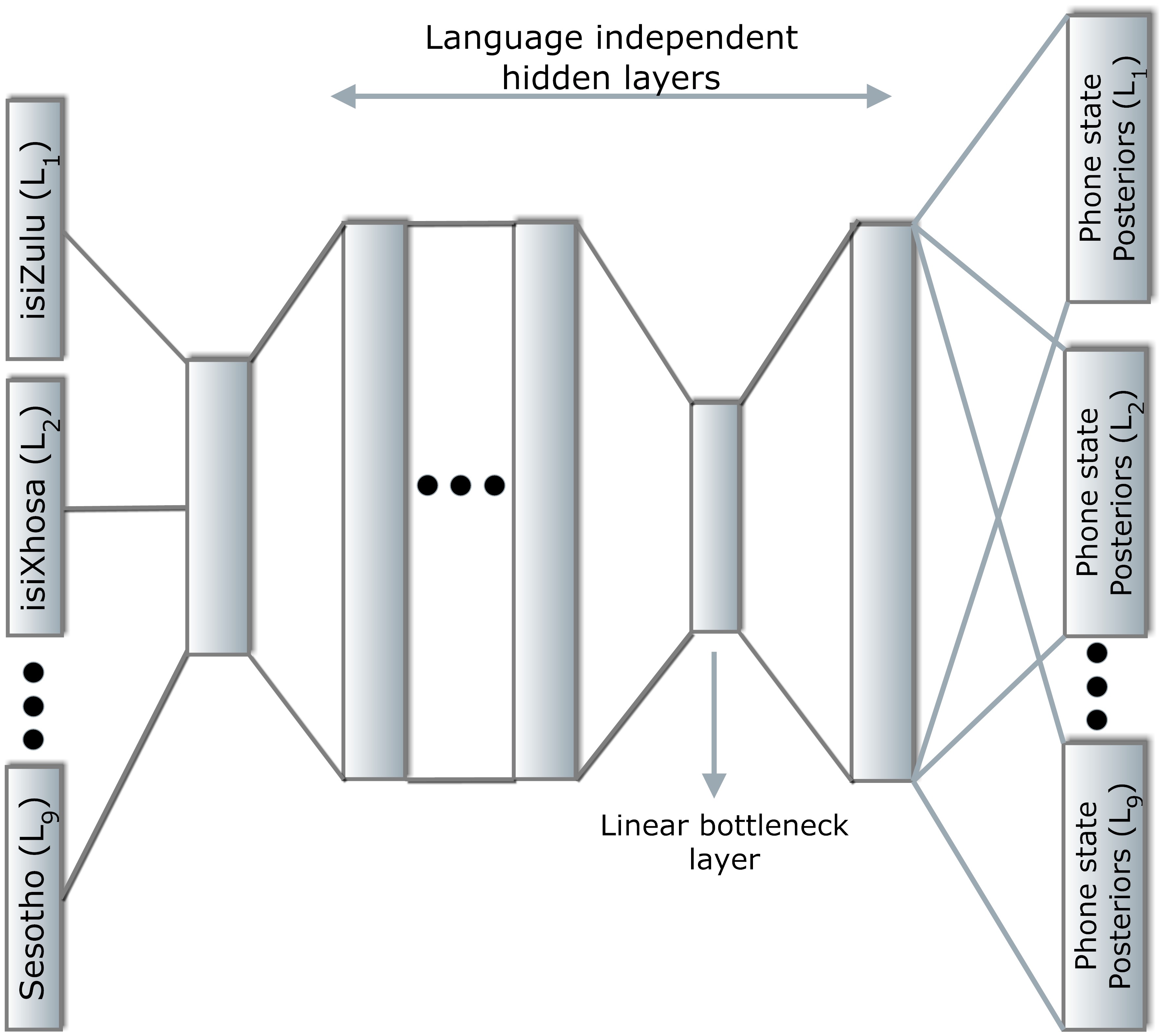}
	\caption{Multilingual bottleneck feature extractor trained on nine South African Bantu languages with block softmax.}
	\label{mBNF}
	\vspace{-6mm}
\end{figure}

To the best of our knowledge, this is the first attempt to develop a mBNF extractor trained on Bantu languages.
Our primary objective is to use the mBNF features to enhance the performance of code-switched ASR.
The extractor was developed using the monolingual NCHLT data introduced in Section~\ref{sec2.1} according to the Babel multilang recipe in the Kaldi ASR toolkit \cite{povey2011kaldi}.
Due to time and computational constraints, fine-tuning of the hyperparameters could not be performed.

As a starting point, a context-dependent Gaussian mixture model-hidden Markov model (GMM-HMM) system is trained for each language to obtain the alignments required for training the feature extraction network.
The features used in this step comprise 39-dimensional MFCCs (including $\Delta$ and $\Delta\Delta$) calculated over a 25ms window size with 10ms overlap between windows.
Three-dimensional pitch features were also included since the Bantu languages are tonal.

A block diagram of the time-delay neural network (TDNN) architecture we used to train the feature extractors is shown in Figure~\ref{mBNF}.
The network is based on the well-established block softmax approach described in \cite{vesely2012language} and \cite{fer2017multilingually}.
It consists of six 1024-dimensional hidden layers followed by a 39-dimensional (mBNF${_1}$) and 80-dimensional (mBNF${_2}$) linear bottleneck layer and terminates in a block softmax output layer.
The hidden layers are shared across languages while the block softmax output layer separates the phone state posterior training targets per language.
The number of output phone state units varies for each block with a minimum of 4520 for Sesotho and a maximum of 4920 for isiZulu.
The input features comprise high resolution 40-dimensional MFCCs (no derivatives), 3-dimensional pitch features and 100-dimensional i-vectors for speaker adaptation.
The bottleneck layer is used for mBNF extraction.

\begin{table*}[h!]
\caption{Word error rate performance of the four bilingual code-switch ASR systems with and without mBNF features.}
\footnotesize
\renewcommand*{\arraystretch}{1.0}
\begin{tabular*}{\textwidth}{@{\extracolsep{\fill}}lcrrrrrrrrrr@{}}
\toprule
\multicolumn{1}{l}{\multirow{2}{*}{\textbf{System}}} & \multirow{2}{*}{\textbf{Feature}} & \multicolumn{2}{c}{\textbf{Avg}} & \multicolumn{2}{c}{\textbf{EZ}} & \multicolumn{2}{c}{\textbf{EX}} & \multicolumn{2}{c}{\textbf{ES}} & \multicolumn{2}{c}{\textbf{ET}} \\
\cmidrule{3-4} \cmidrule{5-6} \cmidrule{7-8} \cmidrule{9-10}\cmidrule{11-12}
\multicolumn{1}{c}{} & \textbf{extractor} & \textbf{Dev} & \textbf{Test} & \textbf{Dev} & \textbf{Test} & \textbf{Dev} & \textbf{Test} & \textbf{Dev} & \textbf{Test} & \textbf{Dev} & \textbf{Test} \\
\midrule
{A} & Baseline MFCC \cite{biswas2020SLTU} & 39.5 & 42.1 & 33.3 & 38.9 & 34.7 & 42.3 & 49.1 & 47.9 & 40.8 & 39.3 \\
{B} & mBNF$_1$ & 38.9 & 41.5 & 32.5 & 38.8 & 34.1 & 41.5 & 49.3 & 46.5 & 39.8 & 39.0 \\
{C}& mBNF$_2$ & 39.4 & 41.0 & 34.0 & 38.0 & 34.0 & 41.1 & 48.8 & 46.7 & 40.8 & 38.1 \\
\bottomrule
\end{tabular*}
\vspace{-6pt}
\label{tab:res1}
\end{table*}

\begin{table*}[h!]
\caption{Language specific WER (\%) (lowest is best)  for English (\textbf{E}), isiZulu (\textbf{Z}), isiXhosa (\textbf{X}), Sesotho (\textbf{S}), Setswana (\textbf{T}) and code-switched (CS) bigram correct (\textbf{Bi$_{\textbf{CS}}$}) (\%) (highest is best) for the test set.}
\footnotesize
\renewcommand*{\arraystretch}{1.0}
\begin{tabular*}{\textwidth}{@{\extracolsep{\fill}}lrrrrrrrrrrrr@{}}
\toprule
\multirow{2}{*}{\textbf{System}} & \multicolumn{3}{c}{\textbf{English-isiZulu}} & \multicolumn{3}{c}{\textbf{English-isiXhosa}} & \multicolumn{3}{c}{\textbf{English-Sesotho}}& \multicolumn{3}{c}{\textbf{English-Setswana}}\\
\cmidrule{2-4} \cmidrule{5-7}\cmidrule{8-10}\cmidrule{11-13}
 &  \multicolumn{1}{l}{\textbf{E}} & \multicolumn{1}{l}{\textbf{Z}} & \multicolumn{1}{l}{\textbf{Bi$_{\textbf{CS}}$}} & \multicolumn{1}{l}{\textbf{E}} & \multicolumn{1}{l}{\textbf{X}} & \multicolumn{1}{l}{\textbf{Bi$_{\textbf{CS}}$}} & \multicolumn{1}{l}{\textbf{E}} & \multicolumn{1}{l}{\textbf{S}} & \multicolumn{1}{l}{\textbf{Bi$_{\textbf{CS}}$}} & \multicolumn{1}{l}{\textbf{E}} & \multicolumn{1}{l}{\textbf{T}} & \multicolumn{1}{l}{\textbf{Bi$_{\textbf{CS}}$}} \\
\midrule
{A} (baseline) & 32.4 & 43.9 & 38.6 & 35.1 & 47.9 & 32.4 & 34.7 & 57.0 & 34.0 & 27.4 & 45.4 & 42.5 \\
{B} & 32.2 & 43.5 & 38.7 & 34.9 & 45.3 & 34.2 & 34.8 & 54.2 & 34.9 & 26.9 & 45.5 & 42.8 \\
{C} & 31.7 & 43.0 & 39.2 & 34.6 & 44.3 & 34.5 & 35.0 & 53.9 & 34.9 & 26.7 & 44.9 & 43.1 \\ 
\bottomrule
\end{tabular*}
\vspace{-6pt}
\label{tab:results2}
\end{table*}

\vspace{-1mm}
\section{ASR for code-switched Speech}
\subsection{Acoustic Model}
All acoustic models were trained using the Kaldi ASR toolkit \cite{povey2011kaldi} and the training data described in Section~\ref{SEC:corpus}. Three-fold data augmentation was applied prior to feature extraction~\cite{ko2015audio}.
The feature set included standard 40-dimensional MFCCs (no derivatives), 3-dimensional pitch and 100-dimensional i-vectors. For the mBNF experiments, combination features were created by appending the mBNFs to these features.

The models were trained with lattice-free maximum mutual information objective~\cite{povey2016purely} using the standard Kaldi CNN-TDNN-F~\cite{povey2018} Librispeech recipe (6 CNN layers and 10 time-delay layers followed by a rank reduction layer) and the default hyperparameters. 
All acoustic models have a single shared softmax layer for all languages as, in general, there is more than one target language in a segment.

No phone merging was performed between languages and the acoustic models were all language dependent. For the bilingual experiments, the multilingual acoustic models were adapted to each of the four target language pairs.

\subsection{Language Model}
The EZ, EX, ES, ET vocabularies respectively contain 11\,292, 8\,805, 4\,233, 4\,957 word types and were closed with respect to the train, development and test sets.
The SRILM toolkit was used to train and evaluate all trigram language models~\cite{stolcke2002srilm}.
The EZ, EX, ES and ET development set perplexities are 425.8, 352.9, 151.5, and 213.3 respectively.
The corresponding values for the test set are 601.7, 788.8, 180.5, 224.5.\footnote{A more detailed description of the development of our code-switched language models is provided in~\cite{biswas2019EZsemi}.}

\section{Results and Discussion}
ASR performance was evaluated by measuring word error rate (WER) on the EZ, EX, ES and ET development and test sets described in Table~\ref{tab:manual_dev_tst}.
In Tables~\ref{tab:res1} and \ref{tab:results2}, System A is the baseline MFCC-based system without BNF features, while Systems B and C use features produced by mBNF${_1}$ and mBNF${_2}$, respectively. 

\subsection{BNF Extractor Performance}
As can be observed in Table~\ref{tab:res1}, for almost all the cases in the four bilingual data sets, the ASR performance of Systems B and C improve over the baseline  when the \mbox{mBNFs} are included in the feature set. 
Hence, it can be concluded that including mBNFs in the acoustic model training improves ASR performance for code-switched speech. 
Furthermore, using 80-dimensional BNFs (System C) offers improved test set performance over using 39-dimensional mBNFs (System B) for three of the four language pairs as well as on average.
However, training acoustic models with the higher dimensional mBNF features requires more computational resources.

\subsection{Language Specific WER Analysis}
\label{subsec:lan_specific_wer}
For code-switched ASR, the recognition performance at code-switch points is of particular interest. 
Language specific WERs and code-switched bigram correct (\textbf{Bi$_{\textbf{CS}}$}) values for the different systems are presented in Table \ref{tab:results2}. 
Code-switch bigram correct is defined as the percentage of words correctly recognised immediately after a language switch. 
All values are percentages.

It is interesting to note that mBNFs contributed significantly to reduce the Bantu WER, especially for isiXhosa and Sesotho.
Modest reductions in WER were also obtained for English in most of the language pairs. This may be because the extractor was only trained on Bantu languages. 
However, ultimately the aim would be to have a feature extractor that has generalised well over all data and could be used to extract features for any language equally accurately. Further, the accuracy at the code-switch points is also substantially higher for Systems B and C compared to the baseline (System A). 
Hence, adding mBNF features enhanced system performance at code-switch points.

Although current improvements are modest, we would like to point out that initial experiments on our code-switching data that used a BNF extractor trained on a proprietary data set containing other languages yielded similar improvements.

\vspace{-2mm}
\section{Conclusions}

We studied the potential benefit of using bottleneck features for acoustic modelling of under-resourced code-switched speech in four South African language pairs. 
A new bottleneck feature extractor was developed using the Bantu languages in the freely-available NCHLT Speech corpus. 
Two bottleneck feature extractors producing mBNFs with different dimensionalities were included in the investigation.
Recognition results have shown that including the mBNFs in the acoustic modelling not only improved the overall ASR performance for mixed-language speech, but also contributed to improving performance specifically at code-switch points. Future work will include the addition of English and other languages to the pool of languages for mBNF extractor training, optimization of network hyperparameters and investigating the trade-off between performance and the bottleneck dimension.

The source code for training the multilingual bottleneck feature extractor is available at \url{https://github.com/ewaldvdw/kaldi/tree/mbnf_cs2020/egs/nchlt_multi_bnfs/s5}.

\section{Acknowledgements}

We would like to thank the former Department of Arts \& Culture (DAC) of the South African government for funding this research. 
We are grateful to e.tv and Yula Quinn at Rhythm City, as well as the SABC and Human Stark at Generations: The Legacy, for assistance with data compilation.
We also gratefully acknowledge the support of NVIDIA corporation with the donation GPU equipment used during the course of this research, as well as the support of Council for Scientific and Industrial Research (CSIR), Department of Science and Technology, South Africa for provisioning us the Lengau CHPC cluster for seamlessly conducting our experiments.

\bibliographystyle{IEEEtran}

\bibliography{mybib}

\end{document}